\documentclass[11pt]{kluwer}

\usepackage{graphicx}

\begin{document}
\begin{article}
\begin{opening}
\title{Cyclic evolution and reversal of the solar magnetic field.\\
I. The large-scale magnetic fields}
\author{R.~N.~Ikhsanov and V.~G.~Ivanov}
\runningauthor{R.~N.~Ikhsanov and V.~G.~Ivanov}
\runningtitle{Cyclic evolution and reversal of the solar magnetic field, I}
\institute{Central astronomical observatory at Pulkovo,
Pulkovskoye chaussee 65/1, 196140, Saint-Petersburg, Russia;
E-mail: solar1@gao.spb.ru}
\date{October 15, 2003}
\begin{abstract}
On the base of the solar magnetic field measurements obtained
in Stanford in 1976--2003 the properties of the cyclic evolution of the
large-scale magnetic field are investigated. Some regularities are found
in longitudinal and latitudinal evolution of the magnetic field in
cycles~21, 22 and 23. The cyclic development of the large-scale magnetic
field can be divided into two main phases. The phase~I, which includes a
period approximately from two years before and until three years after the
maximum of the solar cycle, is studied in detail. It is found that before
the reversal of the large-scale magnetic field the neutral line of the
magnetic field in antipodal longitudinal intervals shifts from the equator
to opposite directions in cycles~21 and~22, but not in cycle~23. During
the sign reversal of the large-scale magnetic field in cycles~21 and~22 in
the antipodal longitudinal intervals the magnetic field of opposite
polarity is observed in all latitudes, thereby forming an equatorial
dipole. After the magnetic field reversal a longitudinal oscillation of
the magnetic neutral line with regard to the equator takes place, which
has a period about 2~years and damps to the minimum of the 11-year cycle.
The intervening longitudinal intervals of the large-scale magnetic field
correspond to positions of the active longitudes of sunspot activity, thus
indicating a close connection of the large-scale and the local magnetic
fields. In evolution of the large-scale magnetic field a periodicity with
period $1.23\pm0.16$~year is revealed, which is close to the period found
by helioseismological  methods in variations of the solar rotation near
the tachocline.
\end{abstract}
\keywords{Sun: magnetic fields -- Sun: activity}
\end{opening}

\section{Introduction}

The knowledge of the large-scale magnetic field properties and its
relation with the local fields of less scales is a determinative factor
for comprehension of solar cyclicity mechanism. The term ``large-scale
field'' is used to be referred to the global (or, in other terms,
background) solar magnetic field (Hoeksema \& Scerrer \cite{hoe86},
Makarov \& Sivaraman \cite{mak89}, Obridko \& Shelting \cite{obr92},
Mikhailutsa \cite{mik95}). However, isolation of the large-scale
magnetic field ``in a pure form'' seems to be a complicated (if possible
at all) task. The reason is that on the solar surface the magnetic field
structures of different scales concurrently exist. At first sight, their
distribution seems to be rather chaotic. However, certain regularities can
be found both in their dimensions and locations.

Until 1970 three ``quiet'' scales of photosphere structural formations
were known: a granule, a supergranule and a giant cell. The investigations
of Ikhsanov (\cite{ikh70,ikh75}), made in the late 60s and based upon
complex analysis of data for cycles 18--20 of solar activity, found out
seven ``quiet'' scales from granules to supergiant granules (for the
latter the term ``cells'' can be only conditionally used). The sequence
numbers of the scales and their names are listed in the first two columns
of Table~\ref{tab1} (in brackets names of the scales are given that are
commonly used by now). The third column of the Table includes bounding
scales, on which local formation with greater magnetic field strength
can emerge (these scales are sometimes called ``magnetic scales''). In the
last column the mean dimensions of the formations are presented. It does
not follows that there are no formations of intermediate scales. For
example, formations twice as large as a mesogranule are rather often
observed. However, the photospheric structures exhibit a tendency to group
in these seven scales. Therefore, the process of formation of such
structures is not pure random, but have some regularities, which are
stochastically manifested as the discrete scales. These scales reflect
dynamical processes that take place both on the solar surface and, more
importantly, in the deep layers of the Sun. So, the scale $CG$ was later
found out by Kawaguchi (\cite{kaw80}), $GCG$ (mesogranules), by November
et al. (\cite{nov81}), Oda (\cite{oda84}), $CsG$, by  MacIntosh \& Wilson
(\cite{mac85}), who called it ``intermediate-sized cell'', etc.

\begin{table}
\begin{tabular}{lp{.3\textwidth}p{.3\textwidth}l}
\hline
N
& \multicolumn{1}{c}{Formation:}
& \multicolumn{1}{c}{Bounding formation:}
& \multicolumn{1}{c}{Mean dimension}\\
& \multicolumn{1}{c}{``quiet'' scale (a)}
& \multicolumn{1}{c}{``active'' scale (b)}
& \multicolumn{1}{c}{(km)}\\
\hline
I&   Supergiant granule SgG&\multicolumn{1}{c}{$\times\times\times$}&
 $1\cdot10^6$\\
\hline
II&  Giant granule gG &
 Very large and complex sunspot group &  $3\cdot10^5$\\
\hline
III& Group of supergranules GsG (intermediate-sized cell) &
 Large and medium sunspot groups & $1\cdot10^5$\\
\hline
IV&  Supergranule sG &
 Medium and small sunspot groups &  $3\cdot10^4$\\
\hline
V&   Group of complexes of granules  GCG (mesogranule) &
 Very small sunspot groups and pores & $1\cdot10^4$\\
\hline
VI&   Complex of granules CG (protogranule) &
Small-scale magnetic formations & $3\cdot10^3$\\
VII& Granule G &
\multicolumn{1}{c}{\dots}& $1\cdot10^3$ \\
VIII&\multicolumn{1}{c}{$\times\times\times$}&
\multicolumn{1}{c}{\dots}& $3\cdot10^2$\\
\hline
\end{tabular}
\caption{The structural formations in the solar photosphere (Ikhsanov
\protect\cite{ikh70,ikh75}).} \label{tab1}
\end{table}

These scales, taken separately, are undoubtedly of a strong interest, and
at present a great attention is focused to their investigation. However,
another important feature of Table~\ref{tab1} is that it indicates
existing of a certain hierarchical network in the structure of the solar
photosphere. Later a similar hierarchy of formations was proposed by
MacIntosh (\cite{mac92}), but he gave only five of the seven scales of
Table~\ref{tab1} (G, CGG, sG, CsG è gG).

However, taking into account the ``quiet'' scales only, without their
relation with the bounding (``magnetic``) ones, does not give an authentic
picture of the solar surface structure. One of essential features of the
structure is that on the boundary of a cell of some given scale local
magnetic fields can emerge, mainly in a form of sunspot groups and
centres of activity. It follows from Table~\ref{tab1} that the dimensions
of the boundary formations are shifted one step below with respect to the
``quiet'' scales. Hence, these formations do not exceed the dimensions of
the preceding ``quiet'' scale. For example, on the boundary of a
supergranule, if it is not sited on the borderline of greater scales,
pores or very small sunspot groups can emerge, and on the border of
a supergiant cell, along with smaller scales, the largest and the most
complex sunspot groups and the corresponding active centres are used to be
formed. These complex sunspot groups, especially when they emerge at a
joint of supergiant cells, are analogues of active longitudes (Ikhsanov
\cite{ikh73}). They rotate as a solid body (Vitinsky \& Ikhsanov
\cite{vit72}), as distinct from smaller sunspot groups, which are formed
on boundaries of smaller scales and manifest differential rotation. It
follows that the largest scale of Table~\ref{tab1} has a character of
global scale. We should note that there are another important conclusions
that can be derived from Table~\ref{tab1}, but in this paper we are
interested, first of all, in separation of the large-scale magnetic field.

Thus, if to treat the bounding scales with strong magnetic fields as
medium-scale (local) formations, then, according to Table~\ref{tab1},
the large-scale magnetic field is to be carried only by formations with
dimensions that are essentially larger than the dimension of the largest
bounding (``magnetic'') scale, i.e. is to have size close to that of
the supergiant cell, which occupies an area comparable with
the radius of the Sun.

Hence, to a first approximation, all formation on the solar surface, as
well as photospheric magnetic fields, can be divided into three types,
which correspond to the small-scale, medium-scale and large-scale magnetic
fields. The boundary between the first and the second scale is the
supergranule, on border of which pores are formed. Therefore, formations
in the range of scales between the supergranule and the giant cell can be
treated as medium-scale formations in the photosphere. Some questions of
organization of the small-scale formation, their properties and relations
with magnetic fields was discussed in the paper of Ikhsanov  et al.
(\cite{ikh97}). In this paper we shall examine regularities in the
structure and evolution of the large-scale magnetic fields.

\section{The latitudinal distribution of the large-scale magnetic field
and its cyclic evolution}

For study of the solar magnetic field evolution over the 11-year cycle we
use as a starting material the long uniform series of the magnetic field
measurements, which was obtained in Stanford in 1976--2003 on a
magnetograph with 3-minute resolution. Thereby, the data include the
large- and medium-scale magnetic fields, which correspond at least to the
three largest scales of Table~\ref{tab1}. The task to separate the
large-scale magnetic field comes, in the main, to removal of the local
magnetic fields. To do it we used two method. The first one is a simple
averaging with the window size equal to the radius of the Sun. The second
method is based on the fact that the local magnetic field, at least in an
early stage, have a bipolar structure, i.e. their strength lines are
closed. In order to remove such local fields, one can use the magnetic
field distribution, reconstructed on the supposition that the strength
lines of the field on some spherical surface that is sufficiently distant
from the Sun (``source surface'') are radial (Hoeksema \& Scerrer
\cite{hoe86}). Hereafter we shall use the magnetic field distribution on
the source surface with radius equal to 2.5 solar radii. On such a
distance from the photosphere the local magnetic fields fade out.
Therefore, the latter method can isolate open magnetic fields of the
largest scale, whereas the former one allows to account the large-scale
fields both with open and closed strength lines.

\begin{figure}
\centering {\includegraphics[width=0.7\textwidth, bb=25 0 225 285]{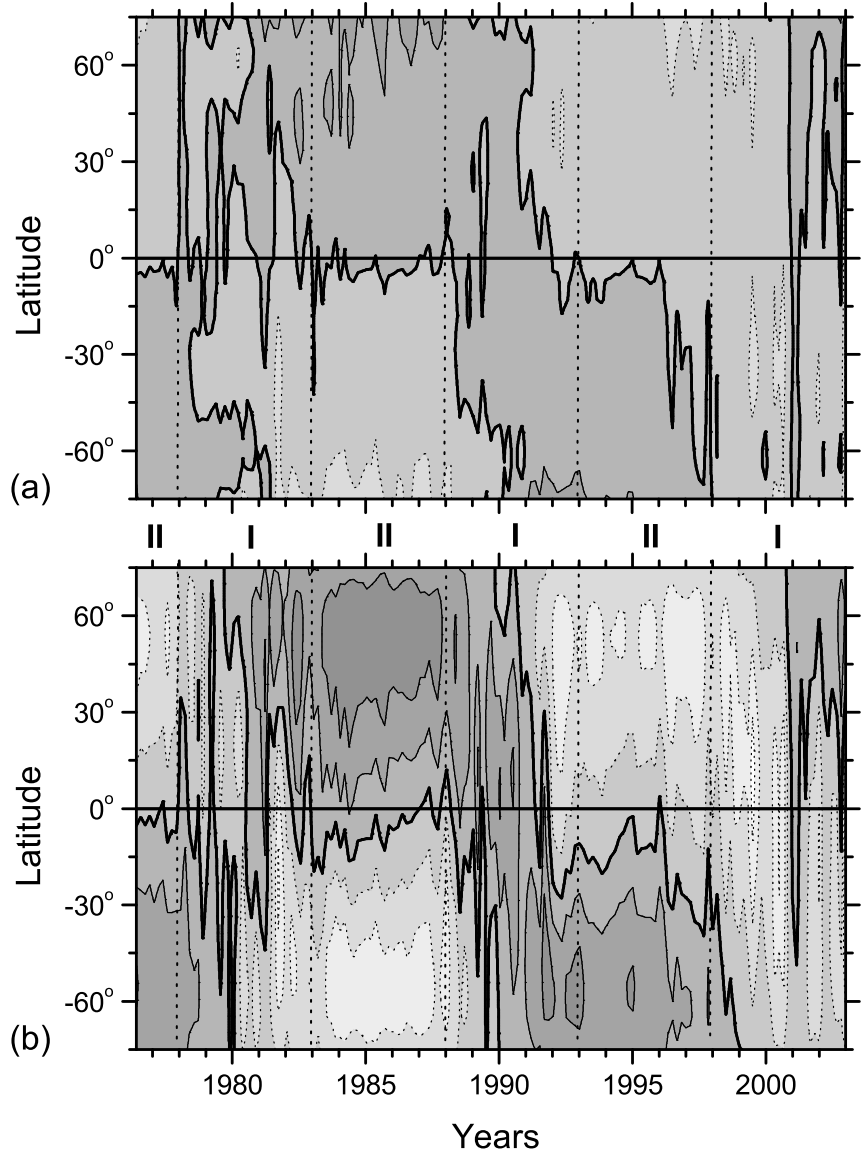}}
\caption{The latitude-time maps of the large-scale solar magnetic field:
the filtered photosphere field with scale $\ge120^\circ$ (a) and the
field on the source surface (b). The light areas correspond to the
positive-polarity magnetic field, the dark areas, to the negative-polarity
one. The borders between the phases~I and~II are marked by
vertical dotted lines.}
\label{fig1}
\end{figure}

The large-scale magnetic fields filtered out by these two methods are
presented in Fig.~\ref{fig1} as latitude-time maps for the time interval
1976--2003. One can see that, being different in some details, these two
maps are as a whole rather similar.

It is evident from Fig.~\ref{fig1}b that two
different phases exist in the 11-year magnetic field evolution. The
phase~I includes the part of the solar cycle approximately from
2~years before its maximum to 2--3 years after it. On this phase
polarity reversals happens frequently, which look like long
latitudinal bands on the maps and cross the equator quite often. It
is especially prominent in low latitudes ($0^\circ{-}\pm30^\circ$).
So in cycle 21 in the north hemisphere the bands of positive polarity
predominate, and in cycle 22, on the contrary, ones of negative
polarity. It is to be noted that the polarity of the north hemisphere
predominates in each of the three investigated cycles. In particular,
it is manifested as ``spilling over'' of the magnetic fields of the
north hemisphere to the south one. The final sign reversal of the
large-scale magnetic field in the 11-year cycle takes place on the
phase~I. This reversal completes consecutively in high, next in
middle, and finally in low latitudes, having total duration about
1.5--2~years. We should note that estimating the moment of the
reversal in high latitudes one must bear in mind that the
measurements were made in latitude range $\pm75^\circ$, thus the
polar reversal can really occur slightly later than it can be
observed on the maps. On the phase~II in all latitudes of either
hemisphere the magnetic field of the uniform polarity is observed,
with the maximum of the field strength being in heliolatitudes
$40^\circ{-}60^\circ$. As approaching to the equator, the field
strength decreases smoothly. In some moments the polarity of the
north hemisphere crosses the equator and intrudes into the south one
up to latitudes $-10^\circ$ and farther.

The above mentioned regularities can be seen as well in Fig.~\ref{fig1}b,
which was obtained by the photospheric magnetic field averaging with the
window size $120^\circ$. It is important that on the phase~I during
2--3~years from its beginning and until the magnetic field reversal the
areas of uniform polarity with the latitudinal dimension
$25^\circ-30^\circ$ are observed in Fig.~\ref{fig1}b, as distinct from the
latitudinal bands in Fig.~\ref{fig1}a. In either of the hemispheres there
are three such areas of alternating polarity. It can be interpreted as
existing of regions of the large-scale magnetic field with closed strength
lines before the cycle maximum,

The division of the phase~I into two epoches, before and after the
magnetic field reversal, will be discussed later. We also shall discuss it
in more detail in the next paper.

\section{Longitudinal and latitudinal evolution of the
large-scale magnetic field in cycles~21--22}

One of the most important problems of investigation of the large-scale
solar magnetic field is looking for the cause of the longitudinal
non-uniformity, which first was found in sunspot distribution, and later,
in other indices of the solar activity (see, e.g., Vitinsky et al.
\cite{vitkop97}, Vitinsky \cite{vit97}, Ivanov \& Ikhsanov \cite{iva98}).

\begin{figure}
\centering {\includegraphics[width=0.7\textwidth, bb=15 0 235 310]{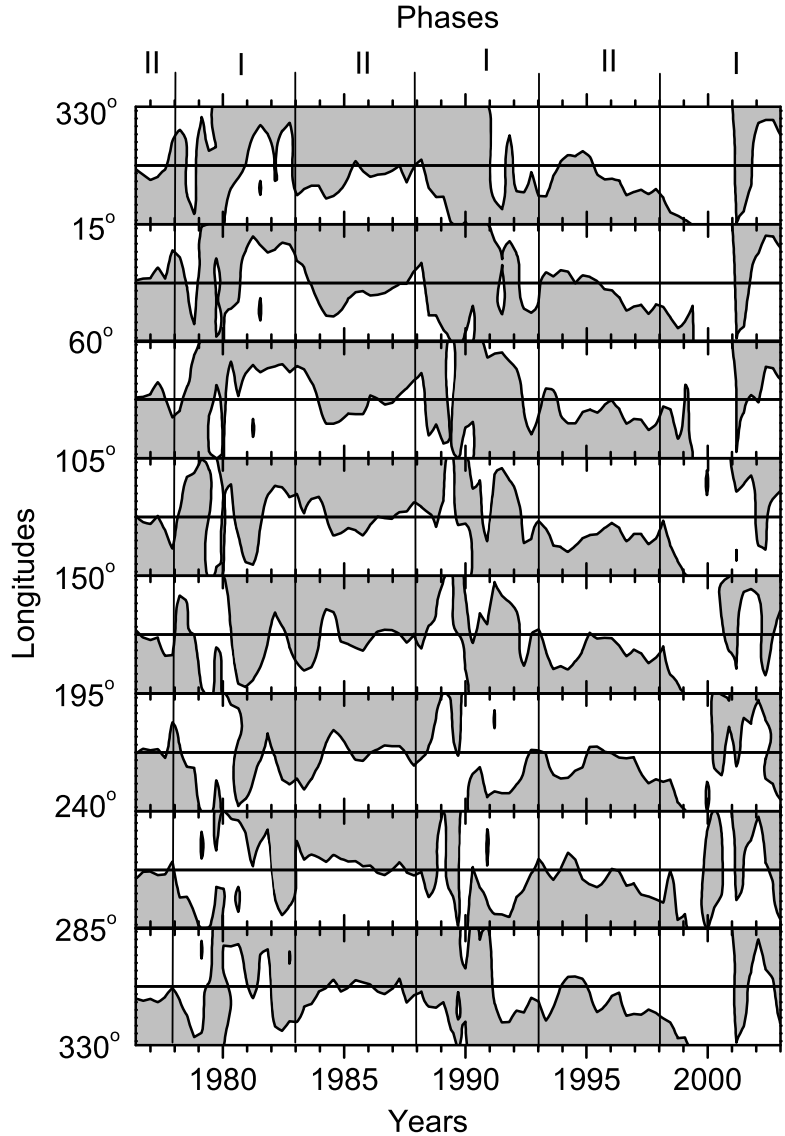}}
\caption{The latitude-time maps of evolution of the magnetic neutral line
in 45-degree longitudinal intervals. The ordinate axis of every map
corresponds to latitude, the labels to the left of the maps indicate the
longitudinal ranges of the intervals.}
\label{fig2}
\end{figure}

To study the longitudinal and latitudinal variations of the
large-scale magnetic field we divided the solar surface into 30-,
45- and 90-degree longitudinal intervals and constructed for every
interval and for time range 1976--2003 a latitude-time magnetic field
distribution. In Fig.~\ref{fig2} the case of the eight 45-degree
intervals and the time step of 1/3 year is presented. First of all,
it follows from Fig.~\ref{fig2} that the picture of evolution of the
magnetic field neutral line varies from one interval to another. In
the distribution of the positive- (light areas) and negative-polarity
(dark areas) magnetic field some regularities can be observed.

So, on the phase~II one can see oscillations of the neutral line. The
oscillations in the different longitudes are not synchronous, being
sometimes in antiphase (e.g., in 1984 or  1993--1994). However,
these oscillations occur only in low latitudes. Quite different picture is
observed on the phase~I, which occupies time interval of the ascending,
the maximum and the beginning of descending of the solar activity,
including the epoch of the magnetic field reversal. In this period the
boundary line of the polarities changes, moving to high latitudes. Thus
the phase~II demands a more detailed study.

\begin{figure}
\centering {\includegraphics[height=0.8\textheight, bb=10 0 215 310]{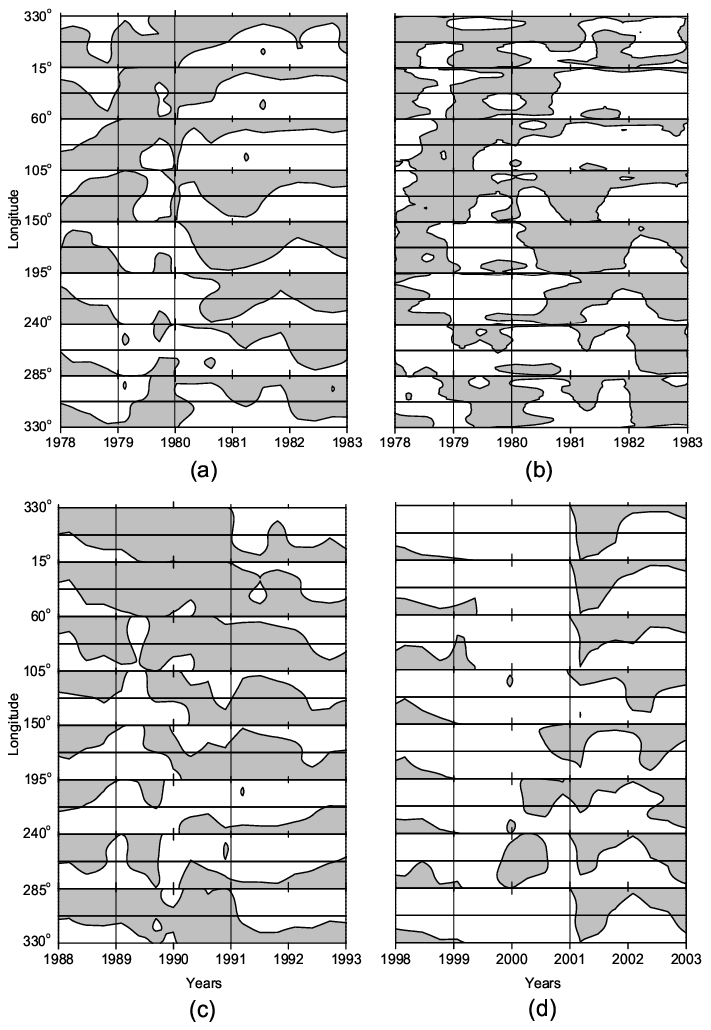}}
\caption{The latitude-time maps of the large-scale
magnetic field in 45-degree longitudinal intervals: in cycle~21 on the
source surface (a) and on the photosphere (b), in cycles~22 (c) and 23 (d)
on the source surface.}
\label{fig3}
\end{figure}

In Fig.~\ref{fig3} the magnetic field evolution on the source surface for
cycle~21 is shown with more details in time. In each of the 45-degree
intervals the field reversal in the north and south hemispheres can be
distinctly observed. In the first three intervals ($330^\circ{-}105^\circ$)
the neutral line, after some incursion into the south hemisphere in 1978,
abruptly rises to the north one, and in the first half of 1979 the
negative polarity attains the N-pole region. In the south hemisphere the
negative polarity exists until 1980.0, whereupon the total polar reversal
is observed at both poles.

In the longitudinal range $150^\circ{-}285^\circ$ the process of the sign
reversal looks differently. Here, on the contrary, motion of the north
polarity into the south hemisphere is observed, and the sign reversal in
the south hemisphere takes place one year earlier than in the north one.

In the ranges $105^\circ{-}150^\circ$ and $285^\circ{-}330^\circ$ the
picture is intermediate in character, but in these intervals the
motion of the neutral line in the opposite directions can be also
observed.

Therefore, the magnetic field reversal continues during the whole
year~1979. It must be underlined that after the polar reversal the
boundary line between polarities is not stable, oscillating with regard to
the equator. One can see that in the 135-degree intervals mentioned above
the latitudinal oscillations initially take place with larger amplitude
and with phase that is opposite to the phase of the oscillations during
the polar reversal. This antiphase behaviour is especially evident in the
longitudinal intervals $15^\circ{-}60^\circ$ and $195^\circ{-}240^\circ$,
which are 180 degrees away one from another. Note, that in the first range
the magnetic field has a negative polarity in $1980\pm1$~year, while in
the second one, on the contrary, a positive polarity.

Hence, the magnetic field reversal in cycle~21 continues
approximately one year. In the first longitudinal interval
($330^\circ{-}105^\circ$) the reversal manifests itself in transition
of the negative polarity from the north to south hemisphere, and in
the second interval ($150^\circ{-}285^\circ$), the positive-polarity
field transfers in the opposite direction. During approximately half
a year in either hemisphere a uniform polarity exists. We should mark
that this regularity is manifested in all eight longitudinal
intervals, with the magnetic field having an opposite sign in the two
described above 135-degree intervals.

To the right from Fig.~\ref{fig3}a the similar Fig.~\ref{fig3}b is shown,
which was obtained by the second method, i.e. by simple averaging of the
photospheric magnetic field. One can see that, in spite of some fuzziness
of the picture (which is especially well seen in high latitudes), the
figure reveals the similar regularities.

From study of the evolution of the magnetic field polarity on the phase~I
of cycle~22 (Fig.~\ref{fig3}c) the similar conclusions can be made, if to
account that the polarity borderline moves in the opposite direction with
regard to cycle~21 and, besides, a remarkable predominance of the
negative-polarity fields are observed in this cycle .

\begin{figure}
\centering
{\includegraphics[width=0.7\textwidth, bb=25 0 230 240]{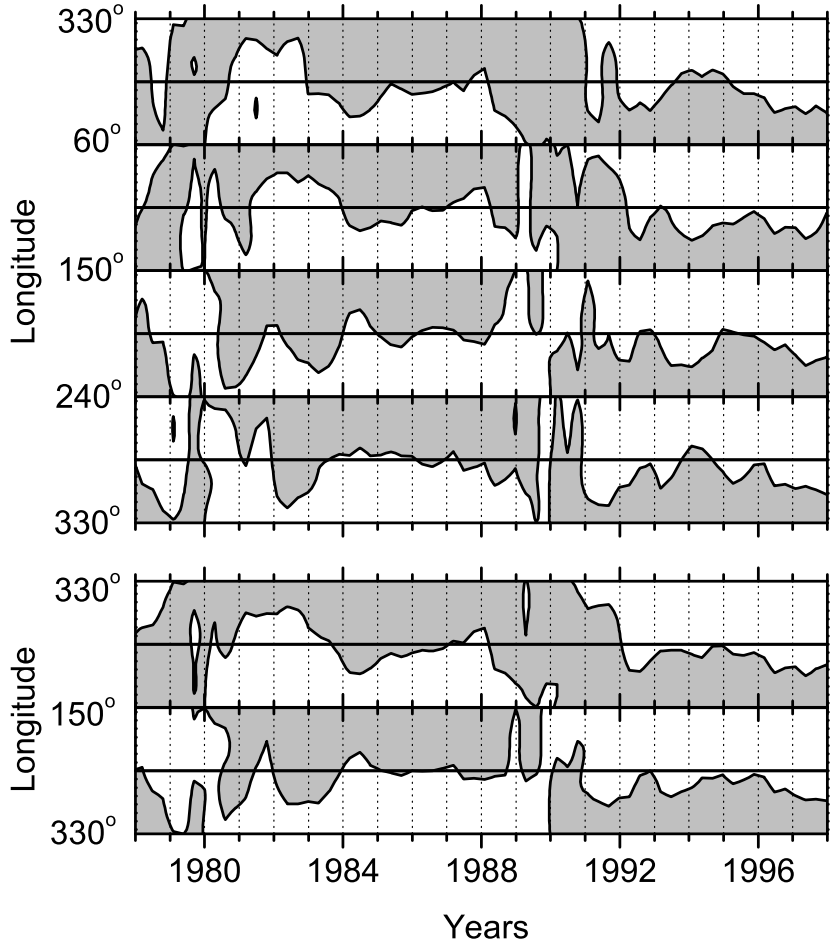}}
\caption{The latitude-time maps of the neutral line evolution in 90-degree
(top) and 180-degree (bottom) longitudinal intervals.}
\label{fig4}
\end{figure}

Combining the successive  45-degree intervals into 90-degree ones
(Fig.~\ref{fig4}), one can see that, both for cycles~21 and~22, in the
first interval the motion of the neutral line is opposite to the third
one, and in the second, to the fourth. Therefore, in the 90-degree
intervals the neutral line on the phase~I evolves in antiphase with the
next nearest interval, rather then with the nearest. This regularity is
evident for cycle~21, and can be also observed for cycle~22, though in the
latter it is noticeably weaker. In every of the 90-degree intervals, both
on the phases~I and~II, the oscillation of the polarity borderline with
regard to the equator are observed (Fig.~\ref{fig4}). During the polar
field reversal the oscillations have maximal amplitudes and spread to all
latitudes up to the poles. Later the amplitude decreases, and on the
phase~II the oscillations take place in low latitudes only. This process
can be interpreted as emerging of a powerful magnetic field impulse during
the polar reversal and its relaxation oscillations of period
1.5--2.5~years, with the large-scale magnetic field of the middle and high
latitudes participating in the process only on the phase~I.

If to examine the picture of the latitude-time magnetic field evolution
after combining of the 90-degree intervals into the 180-degree ones (the
bottom panel of Fig.~\ref{fig4}), it can be seen that the neutral line
oscillations in the west and east hemispheres of the Sun are in antiphase,
forming thereby an equatorial (horizontal) dipole. In the first half of
the phase~I the sign reversal (``flipping'') of the equatorial dipole
takes place. It is especially well manifested in cycle~21.

\begin{figure}
\centering
{\includegraphics[width=0.7\textwidth, bb=25 0 230 305]{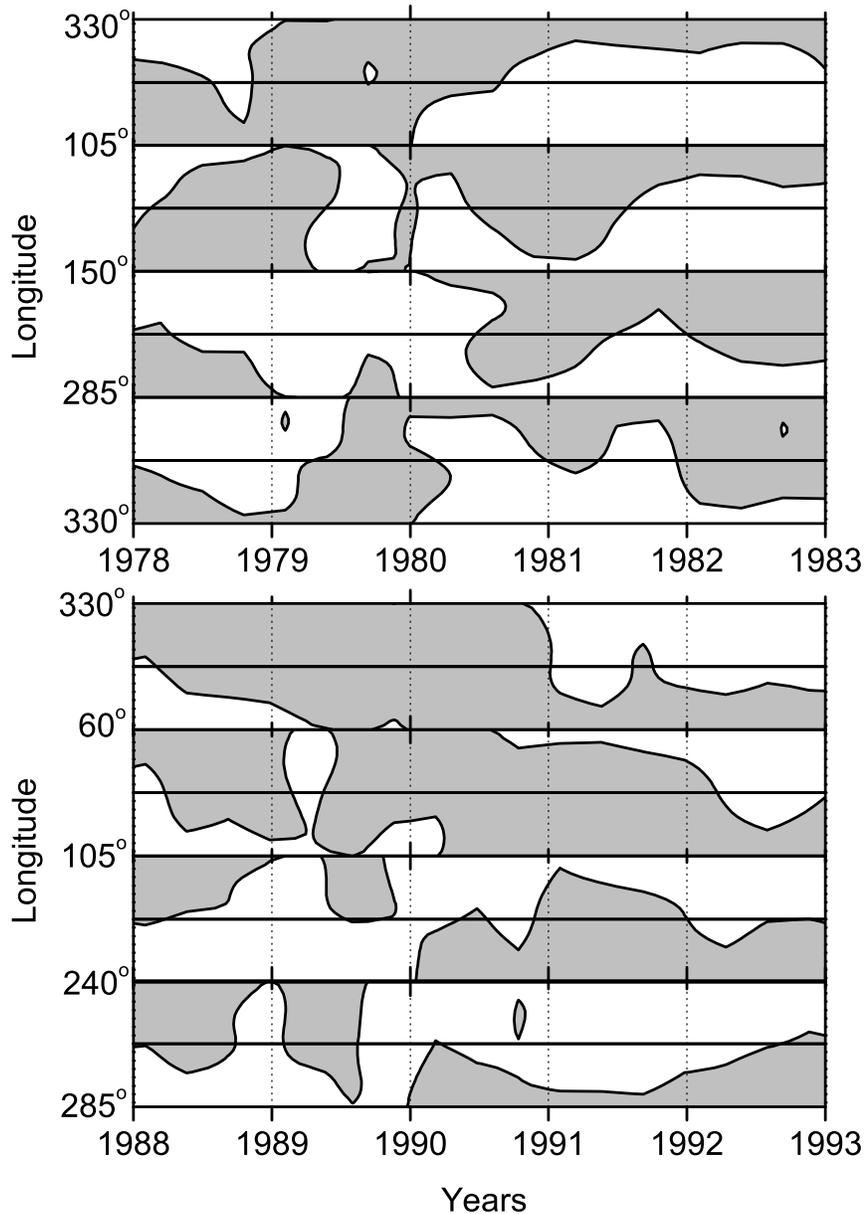}}
\caption{The latitude-time maps of the neutral line evolution in
135-degree and 45-degree longitudinal intervals for the phase~I of
cycles~21 (top) and~22 (bottom).}
\label{fig5}
\end{figure}

More evident antiphase development of the large-scale magnetic field
neutral line in the longitudinal intervals can be observed in two
135-degree and two intervening 45-degree intervals, rather than in
90-degree (as it was marked above in the discussion of Fig.~\ref{fig3}).
In this case (Fig.~\ref{fig5}) the evolution of the horizontal dipole on
the phase~I and its ``flipping'' in the 135-degree intervals with one-year
duration is well revealed, while in the intervening intervals the picture
is more complicated. These intervals, namely $105^\circ{-}150^\circ$ and
$285^\circ{-}330^\circ$ in cycle~21, $60^\circ{-}105^\circ$ and
$240^\circ{-}285^\circ$ in cycle~22 possess a very important feature: they
coincide well with the active longitudes of sunspot activity. According to
estimates of Vitinsky (\cite{vit97}), the most stable active longitudes of
sunspots in cycles~21--23 are sited in intervals $80^\circ-120^\circ$ and
$280^\circ-320^\circ$. The similar positions, $90^\circ$ and $270^\circ$
correspondingly, were obtained by Mordvinov \& Plusnina (\cite{mor01}),
which fairly agrees with places of the above obtained transitional
intervals for the large-scale magnetic field evolution, if to account that
the accuracy of the longitudinal intervals localization is about
$\pm25^\circ$. Such an agreement between the intervening intervals and the
active longitudes indicates a close relation of the global and local
magnetic fields.

\section{Properties of the large-scale magnetic field evolution
in cycles~21--23}

\begin{figure}
\centering
{\includegraphics[width=\textwidth, bb=5 60 620 395]{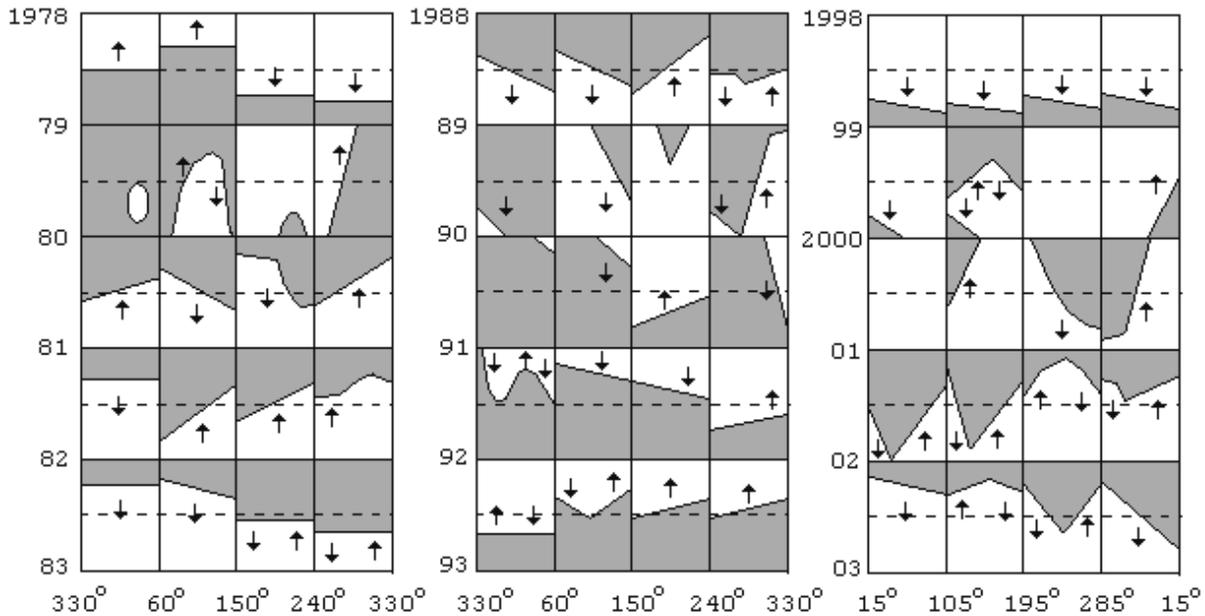}}
\caption{The magnetic neutral line evolution on the phase~I of
cycles~21--23. Every small rectangle is a fragment of the latitude-time
map for 90-degree longitudinal interval and one-year time range. The
values to the left of the maps correspond to years, ones to the bottom
mark the longitudinal ranges. The arrows show directions of the neutral
line motion.}
\label{fig6}
\end{figure}

In Fig.~\ref{fig3} one can readily see a difference between evolution
of the large-scale  magnetic field in cycles~21 and~23. First, in the
21th (and 22nd) cycles before the magnetic field reversal in the
above mentioned antipodal (shifted by $\sim180^\circ$) 45-degree
longitudinal intervals the neutral lines moves from the equator in
opposite directions, while in cycle~23 in all longitudinal intervals
the shift of the neutral line is directed to the south pole of the
Sun. As a result, in the middle of 1999 in almost all longitudes the
large-scale magnetic field of mainly positive polarity was observed.
The reversal of polarity in all longitudes finished as late as in the
beginning of 2001, i.e. it prolonged up to almost two years as
compared with one year in cycle~21. By the way, the anomaly in
evolution of the cycle~23 is observed in the sunspot component as
well. In particular, one of the Gnevyshev-Ohl rules was violated in
this cycle (Vitinsky et al. \cite{vitkop97}). The rule states that
the height of an odd cycle maximum must exceed one of the previous
even cycle. In cycle~23 the maximum of sunspot numbers proved to be
lower than in cycle~22 (121 and 156 correspondingly). What is the
reason of this anomaly? The analysis of the large-scale evolution
shows that it is, possibly, caused by some peculiarities of the
large-scale field development in cycle 22. In particular, in cycle~21
the polarity reversal on all longitudes was observed during
approximately one year, while both in cycles~22 and 23 it lasted
almost two years. Besides, in cycle~21 the areas of the solar surface
occupied by a certain polarity in the epoch of the magnetic field
reversal were approximately equal, while in cycle~22 in the second
year of the reversal (1990) the negative polarity notably predominated. It is
demonstrated by Fig.~\ref{fig6}, where the large-scale solar magnetic
field evolution in the 90-degree longitudinal intervals on the
phase~I is sketched. The arrows indicate direction of the neutral
line motion. In cycle~21 before and during the magnetic field
reversal the opposite motion of the neutral line in the next nearest
90-degree intervals is evident (in 1978), and in~1979 the field in
these intervals has supplementary signs. In cycle~22 this regularity
is violated (in 1990), while in cycle~23 it completely disappears.
Besides, during the field reversal in the pair of cycles~22--23, in
contrast to the pair 21--22, in the most part of the 90-degree
intervals the field polarity reversal is evidently observed.

Another possible reason of such a behaviour of the large-scale magnetic
field is that in cycle~23, as distinct from the 22nd one, an abrupt and
early (in 1992) decreasing of the solar activity was observed. Possibly,
it accounts for the fast motion of the negative-polarity magnetic field to
the south pole after 1996. To the contrary, both in cycles~21 and~22 to
the beginning of the phase~I the neutral line oscillated in the region of
the equator (see Fig.~\ref{fig2}). Therefore, in cycle~23 during the
magnetic field reversal an evident deficit of the negative-polarity
magnetic field was observed. Thus, one can say that in cycle~22, and
particularly in cycle~23, an essential reorganization of the large scale
magnetic field, as compared with the 21st cycle, took place.

Therefore, we can state that the essential changes in the large-scale
magnetic field evolution began in 22nd, i.e. even, cycle, thus confirming
one of the Gnevyshev-Ohl rules.

We should also remark that there is a possibility that the Stanford data
for cycle~23 have some error in position of the zero point for the
magnetic field measurements. This fact can lead to the neutral line shift
and dominating of the positive polarity in the 23rd cycle. However, the new
recalibrated Stanford data do not reveal notable deviations from the data
we used.

We also made a comparison of the Stanford data with the data for magnetic
field of the Sun as a star and found a qualitative agreement between
these two methods of solar magnetic field observations.

\section{The large-scale magnetic field oscillation with period
around 1.3~years}

\begin{figure}
\begin{center}
{\includegraphics[width=0.7\textwidth, bb=35 15 260 350]{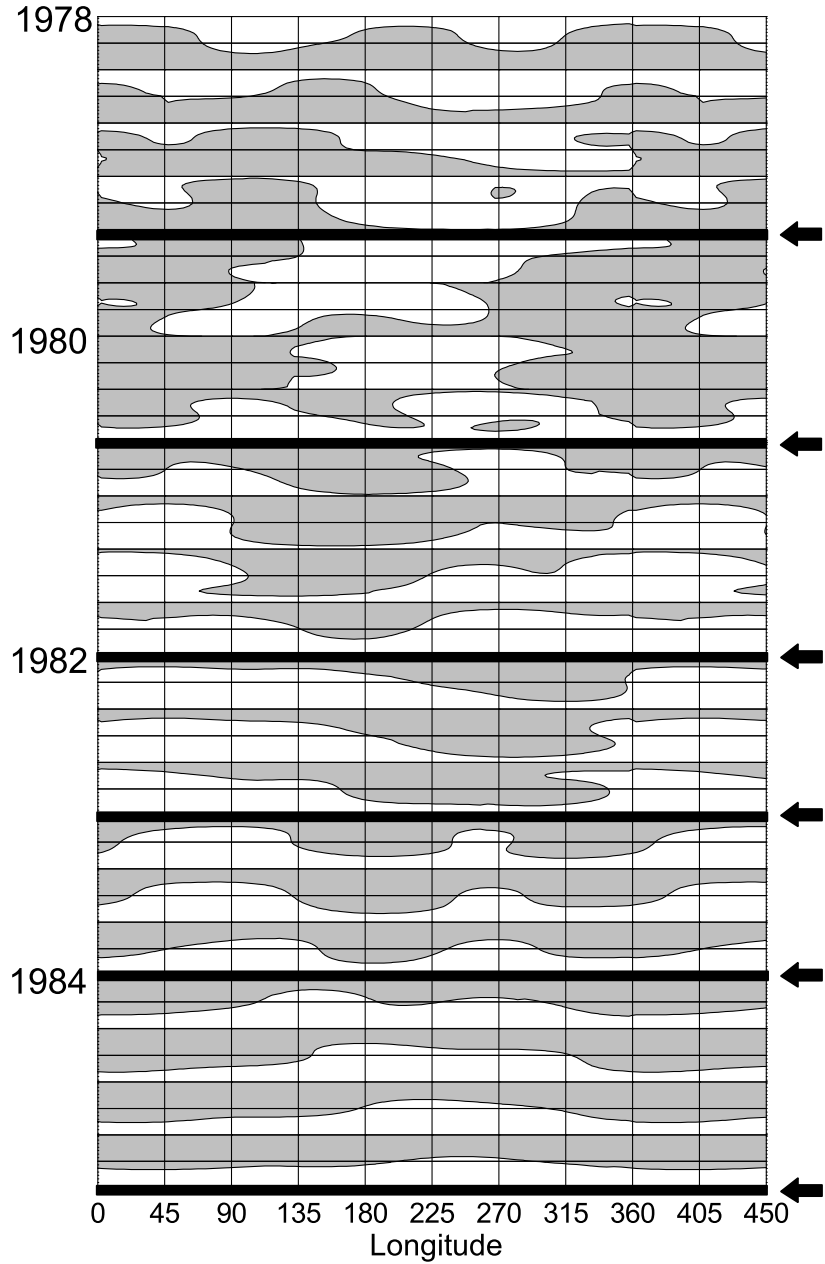}}
\end{center}
\caption{The longitude-latitude maps of the magnetic neutral line position
for successive 4-month intervals in phase~I of cycle~21. The values
to the left mark time, corresponding to the intervals. The bold lines and
arrows to the right indicate boundaries between the different patterns
of the neutral line.}
\label{fig7}
\end{figure}

\begin{figure}
\begin{center}
{\includegraphics[width=0.7\textwidth, bb=35 15 260 300]{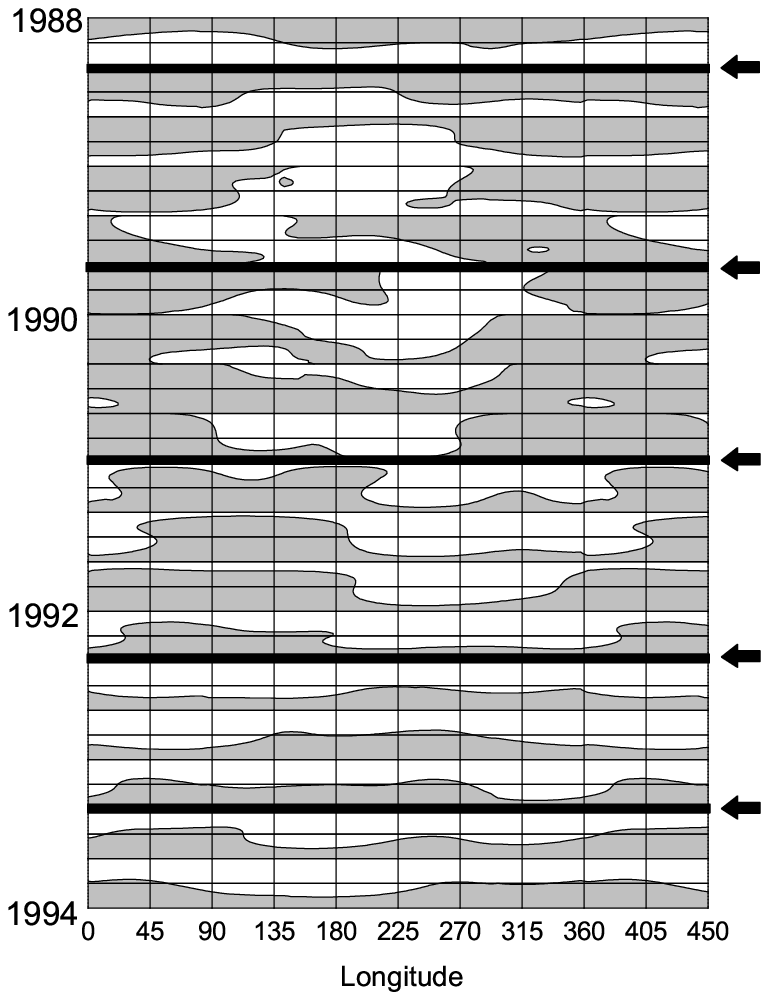}}
\end{center}
\caption{The same as in Fig~\ref{fig7} for the phase~i of cycle~22.}
\label{fig8}
\end{figure}

We have mentioned above that the period between the individual
eruptions of a certain magnetic field polarity in the 45-degree
intervals equals approximately 2 years. But in Fig.~\ref{fig3} one
more periodicity can be observed. In fact, calculation of the time
distance between the successive tops of isogauss in the north and
south hemispheres exhibits recurrence with a period about
$1.22\pm0.13$~years. On the other hand, one can see, particularly on
the plots for cycle~21, that the successive centres of amplitude
maxima of the large-scale magnetic field are separated by the similar
time intervals $\sim1.3$~years. More evidently it can be seen on
another presentation of the magnetic field evolution in
Figs.~\ref{fig7} and \ref{fig8}, where longitude-latitude maps of the
magnetic neutral line position for successive 4-month intervals are
presented. So in cycle~21 (Fig.~\ref{fig7}) the maps are arranged in
groups of some duration (``patterns''), in which the line practically
does not change its longitudinal position. However, on the adjacent
maps with different patterns the neutral line change its longitudes
rather abruptly. The moments of transition of one pattern into
another are marked by the bold arrows in Figs.~\ref{fig7} and \ref{fig8}. With
use of more detailed resolution in time one can reveal that the
transitional period between the patterns is about 2--3 solar
rotations. These patterns are especially evident in epoch after the
reversal, but before this moment they also can be observed rather
confidently, in spite of the fact that the transition from the four-
to two-sector structure of the large scale magnetic field takes place
near this moment.

Every pattern in Figs.~\ref{fig7} and \ref{fig8} is presented by three or
four maps, each corresponding to time interval about 4~months. One can see
that six successive shifts by longitude happens during 7.33~years, i.e.
the process has an oscillation period of about 1.22~year. The longitudinal
magnitude of the shifts are either $90^\circ$ or $180^\circ$. The same picture
can be observed for cycle~22 (Fig.~\ref{fig8}). During five years four
patterns can be observed, thus the periodicity is about 1.25~years, with
the mean for the two cycles being $1.23\pm0.16$~years. One can note that,
in spite of abrupt decreasing of the neutral line oscillation amplitude in
1992, afterwards the patterns can still be observed, though not so
clearly.

Therefore, all three method used above show that in evolution of the
large-scale magnetic field the periodicity of 1.0--1.3~years
($1.23\pm0.16$~years) exists. The period is close in magnitude (and,
possibly, in origin) to the period $\sim1.3$~years, which was observed in
variations of the solar rotation by helioseismological methods. This
periodicity is found in the regions of the tachocline (Howe et al.,
\cite{how00}), which separates the differentially rotating solar
convection zone from more deep region of the Sun with solid rotation. Thus
we obtain an independent confirmation of the statement that the
large-scale magnetic field reflects, to a greater extent, processes taking
place near the lower boundary of the solar convection zone.

Earlier the period of $\sim1.3$~years was discovered by various authors in
different indices of solar activity. For instance, it was found in sunspot
numbers and areas (Kandaurova \cite{kan71}, Akioka et al. \cite{aki87}, in
solar flares (Ishimoto et al. \cite{ish85}, Ikhsanov et al. \cite{ikh88}),
in synoptic $H_\alpha$-charts (Tavastsherna et al. \cite{tav01}).

It was demonstrated by Ikhsanov (\cite{ikh93}), that variation of number
of large sunspot groups with areas more than 1000~m.s.h. in
1980.0--1984.5 reveals a 158--160-day periodicity. It was also found that
the peaks of this oscillation tend to group by three, with the first peak
being the highest, i.e. the triple period of 474--480~days
($\sim1.3$~year) can be observed. The investigated time interval
1980.0--1984.5 corresponds to the epoch when the 150--160-day oscillation
was found in X-ray and gamma-ray radiation of solar flares (Rieger et al.
\cite{rie84}, Dennis \cite{den85}). On the other hand, the identification
of the sunspot groups with the X-ray bursts in this epoch exhibits a clear
relation between the bursts and the largest sunspot groups. According to
Table~\ref{tab1}, the largest sunspot groups concentrate near the boundary
of the largest scale (SgS), thereby manifesting a direct relation with the
large-scale magnetic field.

\section{Conclusions}

It follows from the above analysis of latitudinal and longitudinal
evolution of the large-scale solar magnetic field during three 11-year
cycles (1976--2003~years) that there are certain regularities in their
cyclic variations. Below we list some of them.

\begin{enumerate}
\item
The large-scale magnetic field evolution in 11-year cycle, according to
type of its activity, can be divided into two phases. The phase~I includes
time interval of $\sim$2-3~years near the cycle maximum. In the rest of
the cycle (phase~II) in either hemisphere the magnetic field of a certain
polarity predominates. Besides, the phase~I can be divided into two
subphases, which are separated by the moment of the magnetic field
reversal.

\item
Before the sign reversal in cycles~21 and~22 in all antipodal (i.e.
shifted by $180^\circ$ in longitude) 45-degree intervals the neutral line
moves from the equator in opposite directions.

\item
In cycles~21 and~22 in epoch of the large-scale magnetic field reversal in
the antipodal longitudinal intervals with dimension about $135^\circ$ the
magnetic field of opposite signs is observed in all latitudes. Thus, in
this part of the phase~I the equatorial dipole of the large-scale
magnetic field is evidently manifested.

\item
After the polarity reversal the latitudinal oscillation of the large-scale
magnetic field with regard to the equator is observed, which has a period
1.5--2.5~years and damps to the 11-year cycle minimum.

\item
The transitional longitudinal intervals of the large-scale magnetic field
correspond to the positions of the active longitudes of solar activity,
indicating close relation between the large-scale and the local magnetic
fields.

\item
There are many similar features between cycle~22 and the anomalous
cycle~23 in evolution of their large-scale magnetic field, including
the fact that the regions filled by the negative-polarity magnetic in
the 22nd cycle are filled by the positive-polarity one in the 23rd
cycle.

\item
The period about 1.3~years ($1.23\pm0.16$~years) is found in evolution of
the large-scale magnetic field. The period is close to one found in the
region of the solar tachocline by helioseismological methods. Therefore,
one can assume that the above-mentioned regularities in the large-scale
magnetic field evolution reflect processes, which take place near the
lower boundary of the solar convection zone.

\end{enumerate}

\end{article}
\end{document}